# Human-AI Co-Creation: A Framework for Collaborative Design in Intelligent Systems

Zhangqi Liu

Brown University, Providence, RI 02912 United States, USA

**ABSTRACT**

As artificial intelligence (AI) continues to evolve from a back-end computational tool into an interactive, generative collaborator, its integration into early-stage design processes demands a rethinking of traditional workflows in human-centered design. This paper explores the emergent paradigm of human-AI co-creation, where AI is not merely used for automation or efficiency gains, but actively participates in ideation, visual conceptualization, and decision-making. Specifically, we investigate the use of large language models (LLMs) like GPT-4 and multimodal diffusion models such as Stable Diffusion as creative agents that engage designers in iterative cycles of proposal, critique, and revision.

## INTRODUCTION

The emergence of generative AI technologies has begun to redefine the landscape of design. Traditional computer-aided design tools have largely focused on facilitating efficiency, documentation, and precision. However, the advent of models capable of generating natural language, images, and other creative outputs offers new opportunities for human-AI collaboration that extend beyond productivity. This paper seeks to examine how these generative systems can serve as true co-creators in the design process, particularly in the early ideation stages where ambiguity and exploration are paramount.

The introduction of generative AI into creative workflows marks a departure from conventional paradigms of tool use in design. Historically, designers have operated within frameworks that emphasize mastery over deterministic software environments, tools that respond predictably to user inputs and conform to established design rules. In contrast, generative AI systems such as large language models (LLMs) and diffusion-based image generators introduce elements of surprise, novelty, and ambiguity. These systems do not merely execute commands but produce suggestions that may defy user expectations or diverge from traditional design logic.

This shift presents both opportunities and challenges. On one hand, AI's capacity to generate a broad spectrum of ideas rapidly can help break through creative blockages, stimulate lateral thinking, and accelerate ideation. On the other hand, the stochastic nature of AI outputs may overwhelm users with irrelevant results, introduce biases embedded in training data, or disrupt designers' sense of control and authorship. These tensions highlight the need for new interaction paradigms that emphasize adaptability, interpretability, and co-agency.

Furthermore, the integration of AI into design workflows necessitates a reconsideration of the roles of both human and machine. Rather than viewing AI as a subordinate executor or a mere suggestion engine, we argue for positioning AI as a semi-autonomous collaborator. Such a reconceptualization invites interdisciplinary inquiry: How should designers interact with systems that possess generative agency? What kind of user interfaces best support mutual understanding between human intention and AI inference? What mechanisms are needed to ensure transparency, fairness, and accountability in these co-creative relationships?

This paper addresses these questions by proposing a human-AI co-creation framework that situates generative AI not as a replacement for human creativity, but as a complementary force capable of expanding the bounds of what designers can imagine and produce.

## RELATED WORK

Research on the intersection of artificial intelligence and design has a long lineage, beginning with early rule-based expert systems and evolving through algorithmic design tools, parametric modelling, and adaptive interfaces. In these earlier paradigms, AI systems were often employed to optimize existing design constraints, automate repetitive tasks, or recommend design alternatives based on predefined criteria. These tools, while valuable for enhancing efficiency, lacked the creative capacity or contextual awareness to contribute meaningfully to the ideation phase.

In recent years, advances in machine learning, particularly the emergence of large language models (e.g., GPT series) and multimodal diffusion models (e.g., DALL·E, Stable Diffusion have ushered in a new era of generative tools capable of synthesizing coherent text, visual concepts, and hybrid media representations. Studies have explored the creative potential of these models in domains such as storytelling, graphic design, and architecture, showing that AI can offer non-trivial contributions to the creative process. For example, McCormack et al. (2020) introduce the concept of "computational creativity," emphasizing systems that are not only reactive but capable of intentional novelty. Similarly, recent HCI research has demonstrated how AI can serve as a source of "unexpected inspiration" during brainstorming sessions, helping users overcome functional fixedness.

Despite these promising developments, few studies provide an integrated framework for understanding AI as a co-creative agent across multiple levels of engagement. Much of the current literature remains segmented, focusing on either technical capabilities (e.g., image quality, coherence, semantic alignment) or user experience (e.g., trust, usability, mental model alignment). There is a need for a unifying perspective that bridges these technical and human-centric concerns, addressing how design outcomes and experiences evolve when creativity is distributed between human and machine.

This paper contributes to this gap by synthesizing threads from design theory, AI interaction, cognitive load research, and co-design practices. We build upon foundational concepts such as Boden's (2003) types of creativity (combinational, exploratory, transformational) and extend recent co-creativity frameworks (e.g., Lubart, 2017) to account for real-time, dialogic interactions with generative models.

## METHODOLOGY

To investigate the impact of AI-assisted design tools on creative workflows, we conducted a mixed-methods experimental study involving 24 participants across a spectrum of design expertise. Participants were recruited from both academic and professional environments, including industrial design students, UX/UI practitioners, and freelance visual designers. The selection ensured diversity in experience level, domain familiarity, and creative strategies, which enriched the robustness of our analysis.

Each participant was asked to complete two separate design tasks within a controlled lab environment. The first task employed a conventional design toolset, participants used software platforms such as Adobe XD, Figma, or Sketch to develop a conceptual interface for a hypothetical digital product (e.g., a wellness app or smart home dashboard). The second task introduced AI-assisted tools, wherein participants had access to a prompt-driven large language model (LLM, specifically GPT-4) for ideation, and a multimodal image generation model (Stable Diffusion) to visualize design concepts.

To simulate realistic interaction scenarios, the AI tools were embedded in a prototype co-creation interface that allowed iterative querying, regeneration of outputs, and real-time refinement. For example, designers could enter prompts like "suggest a playful onboarding screen for a fitness app" or "generate a calming dashboard layout for seniors." The system returned text-based ideas and image composites, which users could accept, remix, or discard.

Data collection spanned multiple dimensions. First, we logged all tool interactions, including prompt history, edits, time spent per task, and navigation behavior. Second, we recorded all screen activity and collected the final design artifacts produced under both conditions. Third, we conducted semi-structured interviews immediately following each session, where participants reflected on their cognitive effort, creative confidence, and perceived value of AI suggestions.

To evaluate the cognitive implications of AI assistance, we adapted the NASA-TLX framework for subjective workload assessment and paired it with a fluency scale adapted from Torrance's creativity metrics (e.g., idea quantity, flexibility, elaboration). All qualitative data were transcribed and thematically coded using NVivo, while quantitative data were analyzed via paired t-tests and within-subject ANOVA to assess differences across conditions.

## RESULTS

Our analysis uncovered several statistically and qualitatively significant patterns that illuminate how generative AI impacts design cognition and creative output. The findings are presented in three categories: cognitive effects, creative dynamics, and interactional perceptions.

### Reduction in Cognitive Load

Across all participants, the AI-assisted condition demonstrated a significant decrease in reported cognitive load. NASA-TLX scores were, on average, 22.4% lower compared to the conventional toolset condition ($p < 0.01$). Participants described the AI interface as alleviating "blank canvas anxiety," with one user stating, "I didn't have to stress about where to begin, the AI gave me a jumping-off point." This suggests that AI support serves as a cognitive scaffold, enabling users to bypass initial hesitation and focus on shaping ideas rather than generating them from scratch.

### Enhanced Ideation Fluency and Divergence

Quantitative fluency scores revealed that participants generated a higher number of distinct concepts per minute during the AI-assisted session, with a mean increase of 1.8x compared to baseline. Importantly, these concepts exhibited greater thematic diversity, as judged by blind expert raters assessing originality and variability across designs.

Several participants attributed this to the "provocative unpredictability" of AI outputs. Rather than offering linear extensions of initial ideas, the models frequently introduced unexpected perspectives or stylistic departures. For example, when prompted for a minimalist app layout, one system proposed an organic, nature-inspired interface, a departure that sparked further creative elaboration. These moments of "creative dissonance" appeared to stimulate lateral thinking and exploration beyond habitual design schemas.

| Condition | Cognitive Load (NASA-TLX) ↓ | Ideation Fluency (ideas/min) ↑ | Creativity Rating (1–10) ↑ |
|---|---|---|---|
| Conventional Tools | 75 | 2.1 | 6.4 |
| AI-Assisted Tools | 58 | 3.8 | 8.2 |

Table 1. Quantitative Comparison Between Conditions

Our findings reveal distinct advantages of AI-assisted tools over conventional design environments in terms of cognitive load, ideation fluency, and perceived creativity. Table 1 provides a comparative summary of the experimental results.

### Perception of AI as a Design Partner

Interview data revealed a nuanced shift in how participants perceived the AI system over time. Initially viewed as a passive generator, the AI increasingly took on the role of a "co-pilot" as participants learned to iterate and query with intention. Designers felt more agency when the system provided explanations (e.g., "This layout emphasizes hierarchy by separating task groups visually"), which increased trust and interpretability.

However, not all experiences were positive. Some users reported frustration when AI outputs lacked contextual awareness or failed to follow prompt instructions accurately. There was also ambivalence about ownership; several designers questioned whether the final product could truly be called their own, or if it represented a hybrid authorship.

Participants consistently described AI as a useful catalyst in the early design stages. As one novice designer put it:

*"I usually get stuck staring at a blank screen. But with the AI, even if I didn't like the suggestions, I could react to them. It gave me something to push against."*

Another participant with 10+ years of professional UX experience noted:

*"I expected the AI to be gimmicky, but it actually surprised me with a few out-of-the-box layouts. It felt more like brainstorming with a quirky teammate."*

Some critical reflections also emerged. One designer cautioned:

*"There were moments when I couldn't tell if I was leading the process or just reacting to what the AI generated. It blurred the line between my voice and its influence."*

These qualitative insights reinforce the notion that successful human-AI co-creation depends not only on algorithmic capability, but also on interface design, explainability, and mechanisms for preserving user agency.

These results highlight the importance of transparency, responsiveness, and designer control in shaping effective human-AI co-creation workflows.

**FRAMEWORK FOR HUMAN-AI CO-CREATION**

Drawing from the quantitative and qualitative insights, we propose a three-tiered framework to conceptualize how AI systems can participate in the design process. This framework differentiates modes of collaboration along a spectrum of system initiative and designer control, while also identifying design implications for each level.

**Passive Assistance**

At this foundational level, the AI serves as a reactive tool that responds to user-initiated prompts with static suggestions, e.g., design templates, color palettes, or layout structures. The system does not infer context or user intent beyond surface-level queries. While this mode is low in complexity, it still offers value by accelerating rote decisions and providing starting points.

*Example: A designer types "login page layout" and receives three generic UI mockups. The system does not track revisions or infer user preferences.*

This level benefits novices or time-constrained professionals seeking quick scaffolds but limits deeper creative exploration due to minimal interaction or adaptation.

**Interactive Co-Creation**

In this middle tier, the AI engages in a dialogic process with the designer. Users can iteratively refine prompts, critique outputs, or steer the direction of generation. AI models provide in-situ explanations (e.g., rationale for color harmony, spatial grouping), which helps establish trust and mutual understanding.

*After reviewing a layout, the designer requests "a calmer version for older adults," and the AI not only adjusts spacing and colors, but also explains its reasoning based on accessibility principles.*

This level supports fluent ideation and reduces cognitive strain while keeping the designer in control. It is also where users most often reported perceiving the AI as a collaborative partner rather than a tool.

**Proactive Collaboration**

At the highest tier, the AI system exhibits initiative. It may recognize patterns in the user's style, anticipate needs, or propose radically alternative directions unprompted. While such behavior can spark deep creativity, it also introduces ambiguity regarding authorship and system transparency.

*The AI analyzes a designer's portfolio and autonomously generates speculative UI directions that deviate from the current design language to stimulate exploration.*

To make this level viable, systems must include opt-in controls, clear attribution mechanisms, and transparency protocols to mitigate user disorientation or ethical concerns.

**Framework Summary**

This framework guides not only the evaluation of existing tools but also the principled design of future co-creative systems that balance AI agency with human intention (Table 2).

| AI Role Level | System Behavior | Designer Role | Risks | Opportunities |
|---|---|---|---|---|
| Passive Assistance | Static suggestions | Select, filter | Repetition, lack of depth | Speed, novice onboarding |
| Interactive Co-Creation | Iterative generation, rationale | Direct, critique, refine | Misalignment, overreliance | Engagement, fluency, trust |
| Proactive Collaboration | Autonomous proposal, initiative | Curate, negotiate, reflect | Authorship ambiguity, opacity | Innovation, serendipity, style evolution |

Table 2. Framework Summary

## CONCLUSION AND FUTURE WORK

This paper explored the integration of generative AI into early-stage design workflows, positioning AI as a co-creative partner rather than a passive tool. Our study demonstrated that AI-assisted design can reduce cognitive load, increase ideation fluency, and inspire novel directions through creative dissonance. We introduced a three-level framework: passive assistance, interactive co-creation, and proactive collaboration to describe varying degrees of human-AI engagement.

Key challenges remain in authorship clarity, explainability, and ethical use. Future work will focus on longitudinal impacts, cross-cultural design contexts, and refining AI systems to be more transparent and responsive to human intent. As AI becomes more embedded in creative tools, the goal is not to replace human designers but to meaningfully augment their imagination and agency.